\documentclass[aps,prl,twocolumn,showpacs,preprintnumbers,amsmath,amssymb,showkeys,citeautoscript,superscriptaddress]{revtex4-1}
\usepackage{pdftexcmds}
\usepackage[version=3]{mhchem}
\usepackage{siunitx}
\usepackage{xcolor}
\usepackage{graphicx}
\usepackage{dcolumn}
\usepackage{bm}
\usepackage{amsmath}
\usepackage{tikz}
\usepackage{pgfplots}

\begin{document}
	
\title{Bismuth-doped \ce{Ga2O3} as candidate for $p$-type transparent conducting material}

\author{Fernando P. Sabino}
\email{fernandopsabino@yahoo.com.br}
\affiliation{Department of Materials Science and Engineering, University of Delaware,
Newark, Delaware 19716, USA}

\author{Xuefen Cai}
\email{caixuefen@csrc.ac.cn}
\affiliation{Beijing Computational Science Research Center, Beijing 100094, China}

\author{Su-Huai Wei}
\email{suhuaiwei@csrc.ac.cn}
\affiliation{Beijing Computational Science Research Center, Beijing 100094, China}

\author{Anderson Janotti}
\email{janotti@udel.edu}
\affiliation{Department of Materials Science and Engineering, University of Delaware,
Newark, Delaware 19716, USA}
	

\begin{abstract}

Gallium oxide (\ce{Ga2O3}) is a wide-band-gap semiconductor promising for UV sensors and
high power transistor applications, with Baliga's figure of merit that far exceeds those
of GaN and SiC, second only to diamond. Engineering its band structure through alloying
will broaden its range of applications.
Using hybrid density functional calculations we study the effects on adding Bi
to \ce{Ga2O3}. While in III-V semiconductors, such as GaAs and InAs, Bi tend to substitute on the pnictide site,
we find that in \ce{Ga2O3}, Bi prefers to substitute on the Ga site, resulting
in dilute \ce{(Ga_{1-$x$}Bi_{$x$})_2O3} alloys with unique electronic structure properties.
Adding a few percent of Bi reduces the band gap of \ce{Ga2O3} by introducing \tabularnewline
an intermediate valence band that is significantly higher in energy than the valence band of the host material.
This intermediate valence band is composed mainly of Bi 6$s$ and O 2$p$ orbitals, and it is
sufficiently high in energy to provide opportunity for $p$-type doping.

\end{abstract}


\maketitle

\ce{Ga2O3} is a promising wide-band-gap semiconductor material for power electronics
\cite{Higashiwaki_013504_2012,Higashiwaki_123511_2013,Higashiwaki_034001_2016,Pearton_011301_2018},
solar-blind UV detectors \cite{Orita_4166_2000,Suzuki_131114_2011}, and sensors,
with capabilities that go beyond existing technologies due to its very large band
gap of about 4.7 eV \cite{Orita_4166_2000}, compared to 3.26 eV in 4H-SiC \cite{Patrick_A1515_1965},
and 3.44 eV in GaN \cite{Madelung_1982}.
The ultra wide gap and large critical electric-field (EC) strength of 8 MV/cm of \ce{Ga2O3}
allow for high temperature and high voltage operation,
placing \ce{Ga2O3} at the top of the most promising semiconductor materials for
electronic power switches, just below diamond according to the Baliga's figure of
merit (BFOM) due to its high electron mobility and breakdown electric field \cite{Baliga_455_1989,Higashiwaki_034001_2016}. 
Engineering its band structure would open new avenues in device applications.

\ce{Ga2O3} can be found in five different polymorphs, $\alpha$ \cite{Yoshioka_346211_2007,Sabino_155206_2014},
$\beta$ \cite{Geller_676_1960,Sabino_155206_2014}, $\gamma$ \cite{Arean_35_2000}, $\delta$ \cite{Roy_719_1952,Playford_2803_2013,Sabino_155206_2014},
and $\epsilon$ \cite{Kroll_3296_2005,Yoshioka_346211_2007,Sabino_155206_2014},
all of which show similar electronic structure: a highly dispersive conduction band
derived mostly from Ga 4$s$ orbitals, giving relatively high
electron mobility at room temperature, and a flat, low lying valence band
derived from O 2$p$ orbitals. These features make \ce{Ga2O3} easy to be dope $n$-type,
for instance with Si, Ge, and Sn substituting on Ga site, or F substituting on O site \cite{Orita_134_2002,Muller_34_2013,Varley_142106_2010,Higashiwaki_034001_2016},
yet rather difficult (or perhaps impossible) to be doped $p$-type \cite{Zhang_383002_2016}.

Acceptor impurities in \ce{Ga2O3} induce deep levels in the gap, with rather high ionization
energies. For example, predicted acceptor ionization energies of \ce{Mg}, \ce{Cd}, \ce{Zn}
or \ce{N} in $\beta$-\ce{Ga2O3} are higher than \SI{1}{\electronvolt} \cite{Lyons_05LT02_2018},
so that these impurities will not be activated at typical device operating temperatures.
Moreover, holes in the valence band of \ce{Ga2O3} tend to localize on individual
O atoms, forming self-trapped holes or small hole polarons \cite{Varley_081109_2012},
giving rise to a broad photoluminescence peak well below the fundamental band gap \cite{Yamaoka_93_2016,Ho_115163_2018,Yamaoka_012030_2019}.

A possible way to overcome the deep acceptor levels and the self-trapping hole in \ce{Ga2O3},
thus enabling $p$-type conductivity, is to raise its valence band. This in principle could be
achieved by adding \ce{S} or \ce{Se}, whose valence $p$ orbitals are higher in energy and much
more delocalized than the \ce{O} 2$p$ orbitals. Thus, \ce{S} or \ce{Se} substituting
on the O sites would lift the valence band of \ce{Ga2O3}\cite{Hiramatsu_125_2002,Hiramatsu_012104_2007},
facilitating $p$-type doping.
However, solubility of chalcogenides on the O site is extremely low, in large part due to
the very large atomic size mismatch between \ce{S} or \ce{Se} and O. Alternatively,
one could increase the covalent character of the top of the valence band through
a hybridization of \ce{O} $p$ and metal lone-pair $s$ orbitals \cite{Kawazoe_939_1997,Zhang_383002_2016}.
For example, it has been demonstrated that post transition metal additions, for instance,
\ce{Sn} in \ce{SnO}, \ce{Bi} in \ce{Ba2BiTaO6} or \ce{Bi} in \ce{In2O3} \cite{Ogo_2187_2009,Yabuta_072111_2010,Bhatia_30_2016,Sabino_034605_2019}, leads to higher and more delocalized valence bands.

\ce{Bi} is known to incorporate in the pnictide site in III-V semiconductors as an isovalent group-V anion, such
as in InGaAs and InSb \cite{Janotti_115203_2002}, or to form compounds
with chalcogenides and O, such as \ce{Bi2Se3}, \ce{Bi2Te3} \cite{Shuk_179_1996},
and \ce{Bi2O3} \cite{Walsh_235104_2006,Matsumoto_094117_2010}, where Bi enters
as a trivalent element. Based on atomic size and valence considerations, we explore adding
Bi to $\beta$-\ce{Ga2O3}, in the form of dilute \ce{(Ga_{1-$x$}Bi_{$x$})2O3} alloys.
We investigate the site preference, the mixing enthalpy, and the effects of
Bi incorporation on the band gap and band-edge positions of the alloys as a function of Bi
concentration. We find that Bi prefers to replace Ga at the octahedral sites, and introduces an intermediate
valence band, which is significantly higher than the original O 2$p$ band in
$\beta$-\ce{Ga2O3}, thus, provides an opportunity for enhanced $p$-type doping when the defect level and the host occupied level is decoupled. We also show that minimum energy transitions from the new intermediate valence band and the conduction band is still higher that the visible light range, suggesting that Bi doped \ce{Ga2O3} is a strong candidate for $p$-type transparent conducting material.


Our electronic structure calculations are based on the density functional theory \cite{Hohenberg_B864_1964,Kohn_A1133_1965}
and the hybrid functional of Heyd-Scuseria-Ernzerhof (HSE)  \cite{Heyd_7274_2004,Heyd_219906_2006}
as implemented in Vienna {\em Ab-initio} Simulation Package (VASP) \cite{Kresse_13115_1993,Kresse_11169_1996}.
The interaction between the valence electrons and the ionic cores are treated
using the projected augmented wave potentials \cite{Blochl_17953_1994,Kresse_1758_1999},
with the valence configuration \ce{O}:$2s^{2}2p^{4}$, \ce{Ga}:$3d^{10}4s^{2}4p^{1}$,
and \ce{Bi}:$5d^{10}6s^{2}6p^{3}$. For structure optimizations, we use the
exchange and correlation functional proposed by Perdew-Burke-Ernzerhof
and parametrized for solids (PBESol) \cite{Perdew_136406_2008}, with a
\SI{620}{\electronvolt} cutoff for the plane wave basis set. For the electronic
band structure and density of states (DOS) calculations, however, we use the HSE hybrid functional with a
\SI{470}{\electronvolt} cutoff.

In the HSE formulation, the exchange functional is separated in long and short
range parts \cite{Heyd_7274_2004,Heyd_219906_2006}, and the Hartree-Fock is mixed
with the PBE exchange only in the short range part. We used a mixing of 32\% Hartree-Fock
exchange, which gives a band gap for $\beta$-\ce{Ga2O3} of \SI{4.70}{\electronvolt},
in good agreement with the reported experimental values \cite{Tippins_A316_1965,Ueda_933_1997,Orita_4166_2000}.

For simulating the \ce{(Ga_{1-$x$}Bi_{$x$})2O3} alloys we use special quasi random structures
(SQS) \cite{Zunger_353_1990,Wei_9622_1990} based on a supercell with 120 atoms. Configurations
with Bi concentrations of 2.08\%, 4.17\%, 6.25\%, and 12.5\% were generated by
replacing 1, 2, 3, and 6 Ga atoms with Bi, respectively. For integrations
over the Brillouin zone, we use a $8{\times}8{\times}4$ $k$-point mesh for the
10-atom primitive cell of $\beta$-\ce{Ga2O3}, and an equivalent $k$-point density
for the 120-atom alloy supercells. The optical properties of $\beta$-\ce{Ga2O3} and \ce{(Ga_{1-$x$}Bi_{$x$})2O3}
alloys are obtained from the calculated frequency-dependent dielectric matrix \cite{Gajdos_045112_2006}
using the Kramer-Kronig relations with small Lorentzian broadening parameter of \SI{5d-4}{\electronvolt}.
Phonon-assisted indirect transitions or excitonic effects are neglected.



\ce{Ga2O3} is most stable in the monoclinic $\beta$ structure
belonging to the $C/2m$ space group, with $10$ atoms per primitive cell (2 formula units).
There are two inequivalent cation sites, one where each \ce{Ga} is bonded to four
\ce{O} forming a perfect tetrahedron, and another where \ce{Ga} is bonded to six \ce{O}
forming a distorted octahedron. Therefore, there are two possibilities for \ce{Bi} substituting \ce{Ga} in
$\beta$-\ce{Ga2O3}. Both configurations were tested, and we find that \ce{Bi} prefers, by far, to replace the \ce{Ga}
in the octahedral site, with a total energy difference of \SI{0.59}{\electronvolt} per Bi between the two
configurations. Note that in  $\alpha$-\ce{Bi2O3} \cite{Malmros_383_1970},
the \ce{Bi} atoms are found in high coordinated site, which is similar to the octahedral site in $\beta$-\ce{Ga2O3}.
Given the large energy difference between the octahedral and tetrahedral configurations, in the following
we consider only \ce{Bi} substituting \ce{Ga} on the octahedral sites.

\begin{table}[t!]
\centering
\caption{Mixing enthalpy ($\Delta H^f$) per cation, lattice parameter
and volume per formula unit for different \ce{Bi} concentrations ($x$) in \ce{(Ga_{1-$x$}Bi_{$x$})_2O3}
alloys.
}
\label{lattice}
\begin{ruledtabular}
\begin{tabular}{llcc}
System                        &~$x$              & $\Delta H^f$            & Volume         \\
                              &($\%$)            & (meV/cation)            & (\AA$^3$/f.u.)  \\
\hline
\ce{Ga2O3}                    & ~$0.00$          & -                       & 52.68        \\
\ce{(Ga_{1-$x$}Bi_{$x$})_2O3} & ~$2.08$          & 30.40                   & 53.39         \\
\ce{(Ga_{1-$x$}Bi_{$x$})_2O3} & ~$4.17$          & 61.24                   & 54.05         \\
\ce{(Ga_{1-$x$}Bi_{$x$})_2O3} & ~$6.25$          & 88.34                   & 54.75          \\
\ce{(Ga_{1-$x$}Bi_{$x$})_2O3} & ~$12.5$          & 144.16                  & 56.98          \\
\end{tabular}
\end{ruledtabular}
\end{table}

\begin{figure}[t!]
\centering
\includegraphics[width=0.9\linewidth]{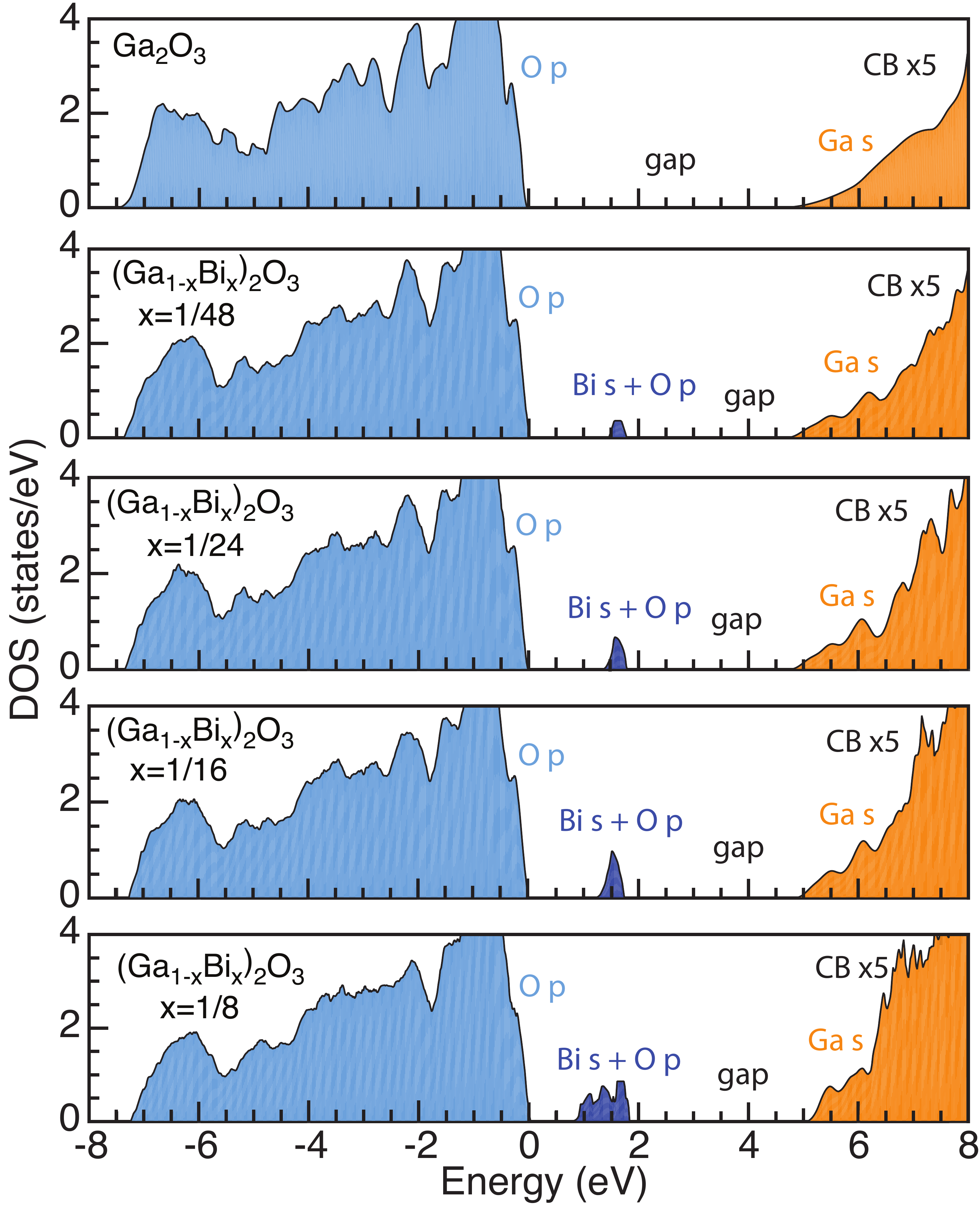}
\caption{
Calculated density of states (DOS) per formula unit showing the effects of \ce{Bi} incorporation in $\beta$-\ce{Ga2O3}.
From top to bottom, \ce{Ga2O3}, and \ce{(Ga_{1-$x$}Bi_{$x$})_2O3} alloys with $x=1/48$, $x=1/24$, $x=1/16$, and $x=1/8$.
The intermediate valence band derived from a hybridization between \ce{Bi} 6$s$ and \ce{O} 2$p$
orbitals are indicated by the dark blue color.
The zero in the energy axis is arbitrary placed on top of the \ce{O} 2$p$ bands, indicated by the light blue color.
}
\label{DOS}
\end{figure}

The calculated lattice parameters for $\beta$-\ce{Ga2O3}, $a_0 = 3.05$ \AA, $b_0 = 12.26$ \AA, $c_0 = 5.81$ \AA, and $\beta = 103.72^{\circ}$,
are in good agreement with experimental data, $a_0^{exp} = 3.04$ \AA,
$b_0^{exp} = 12.23$ \AA, $c_0^{exp} = 5.80$ \AA, and $\beta^{exp} = 103.70^{\circ}$ \cite{Geller_676_1960}.
Adding \ce{Bi} to \ce{Ga2O3} leads to an increase in the lattice parameters due to
the larger \ce{Bi} atomic radius compared to \ce{Ga}. The calculated volume per
formula unit of \ce{Ga2O3} and \ce{(Ga_{1-$x$}Bi_{$x$})_2O3} alloys shows a linear
behavior with the concentration $x$, as listed in Table~\ref{lattice}.

The mixing enthalpy of the \ce{(Ga_{1-$x$}Bi_{$x$})_2O3} alloys are calculated according to the expression:
\begin{eqnarray}
\Delta H^f &=E_{tot}[({\rm Ga}_{1-x}{\rm Bi}_{x})_2{\rm O}_3] \nonumber \\
           &-(1-x)E_{tot}({\rm Ga}_2{\rm O}_3) -xE_{tot}({\rm Bi}_2{\rm O}_3),
\end{eqnarray}
where $E_{tot}[({\rm Ga}_{1-x}{\rm Bi}_{x})_2{\rm O}_3]$ is the total energy of alloy supercell,
$E_{tot}({\rm Ga}_2{\rm O}_3)$ is the total energy of $\beta$-\ce{Ga2O3}, and $E_{tot}({\rm Bi}_2{\rm O}_3)$ is
the total energy of $\alpha$-\ce{Bi2O3}. The calculated mixing enthalpy of the alloys
are also listed in Table~\ref{lattice}.


\begin{figure}[t!]
\centering
\includegraphics[width=0.90\linewidth]{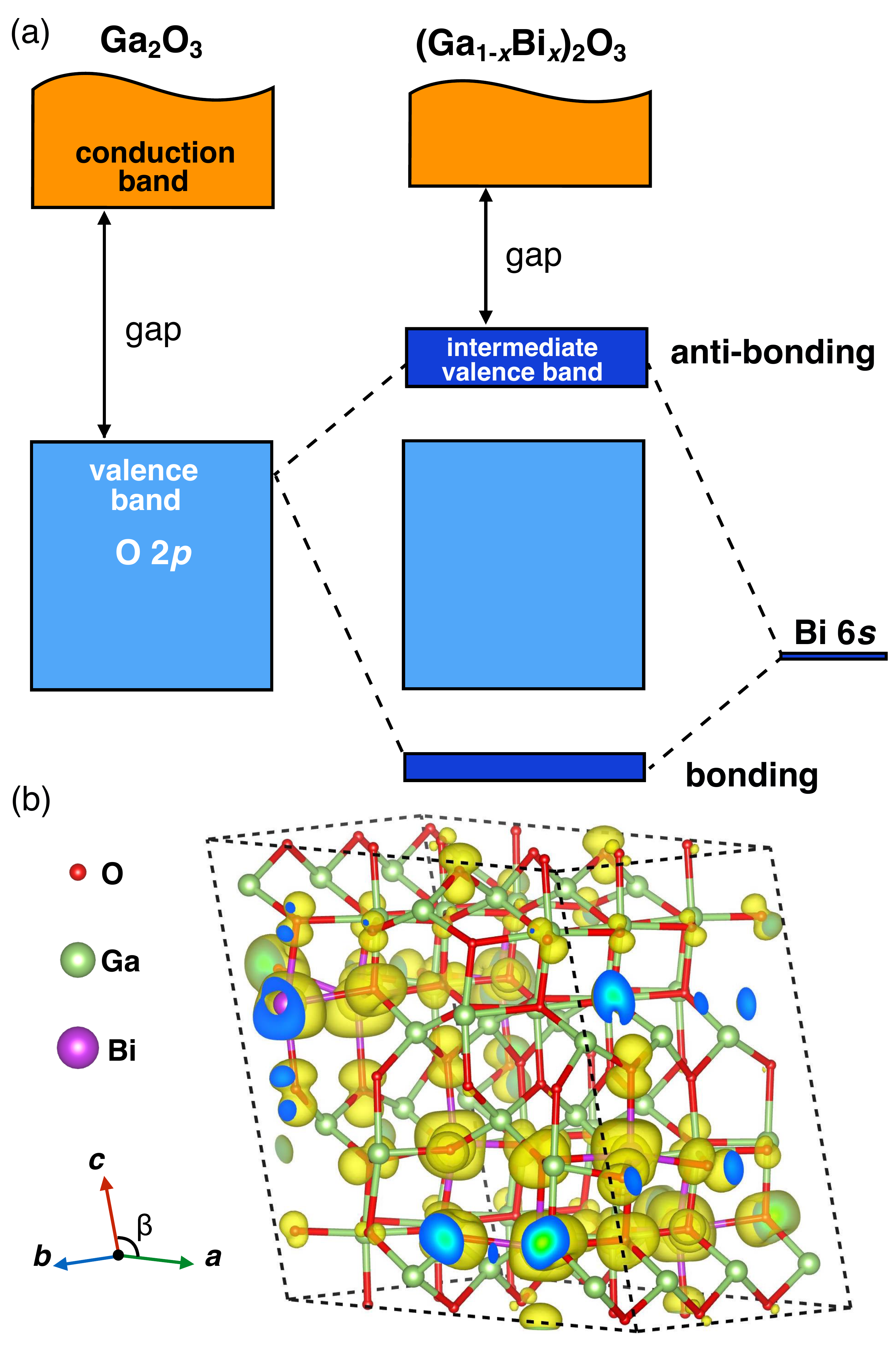}
\caption{
(a) Schematic representation of the coupling between the \ce{Bi} 6$s$ orbitals and the \ce{O} 2$p$ band to
form the occupied intermediate valence band in \ce{(Ga_{1-$x$}Bi_{$x$})_2O3} alloys. (b) Charge
density distribution of the states that compose the intermediate valence band in the
\ce{(Ga_{1-$x$}Bi_{$x$})_2O3} alloy with $x=1/8$. The isosurface corresponds to 5\% of the maximum charge density.
}
\label{coupling}
\end{figure}

For \ce{Bi} concentrations lower than $12.5\%$, the calculated mixing enthalpies
for \ce{(Ga_{1-$x$}Bi_{$x$})_2O3} are not much higher than those of \ce{(In_{1-$x$}Ga_{$x$})_2O3} \cite{Peelaers_085206_2015},
which have been grown experimentally in bulk form \cite{Baldini_552_2014}.
For instance, the mixing enthalpy for \ce{(In_{1-$x$}Ga_{$x$})_2O3} with $x = 25\%$
is approximately 80 meV/cation, which is close to \ce{(Ga_{1-$x$}Bi_{$x$})_2O3} with
$6.25 \%$ of \ce{Bi}. These results suggest that \ce{(Ga_{1-$x$}Bi_{$x$})_2O3} can
in principle also be obtained experimentally, at least for dilute concentrations of \ce{Bi}.

Although the structural parameters are only slightly affected by the incorporation of small concentrations
of \ce{Bi} in \ce{Ga2O3}, we find the effects on the electronic structure to be much more substantial.
In Fig.~\ref{DOS} we show the calculated density of states (DOS) of the
\ce{(Ga_{1-$x$}Bi_{$x$})_2O3} alloys compared to the parent compound \ce{Ga2O3}.
For \ce{Ga2O3}, the states near the valence-band maximum (VBM) have large contribution from the
\ce{O} 2$p$ orbitals, while the states near the conduction-band minimum (CBM) are mostly
derived from the Ga 4$s$ orbitals.
As \ce{Bi} is incorporated into \ce{Ga2O3}, we observe a fully occupied band
significantly higher in energy than the original valence band of the host material \ce{Ga2O3}.
This intermediate valence band comes from the coupling of \ce{Bi} 6$s$, lying well
below the \ce{O} 2$p$ band, and the \ce{O} 2$p$ orbitals forming the higher lying valence bands,
as shown in Fig.~\ref{coupling}(a), featuring an antibonding character. We also note that the charge
distribution associated with the intermediate valence band of the \ce{(Ga_{1-$x$}Bi_{$x$})_2O3}
with $x=1/8$, as shown in Fig.~\ref{coupling}(b), follows the distribution of Bi, which is quite uniformly distributed across the whole crystal.

\begin{figure}[t!]
\centering
\includegraphics[width=1.00\linewidth]{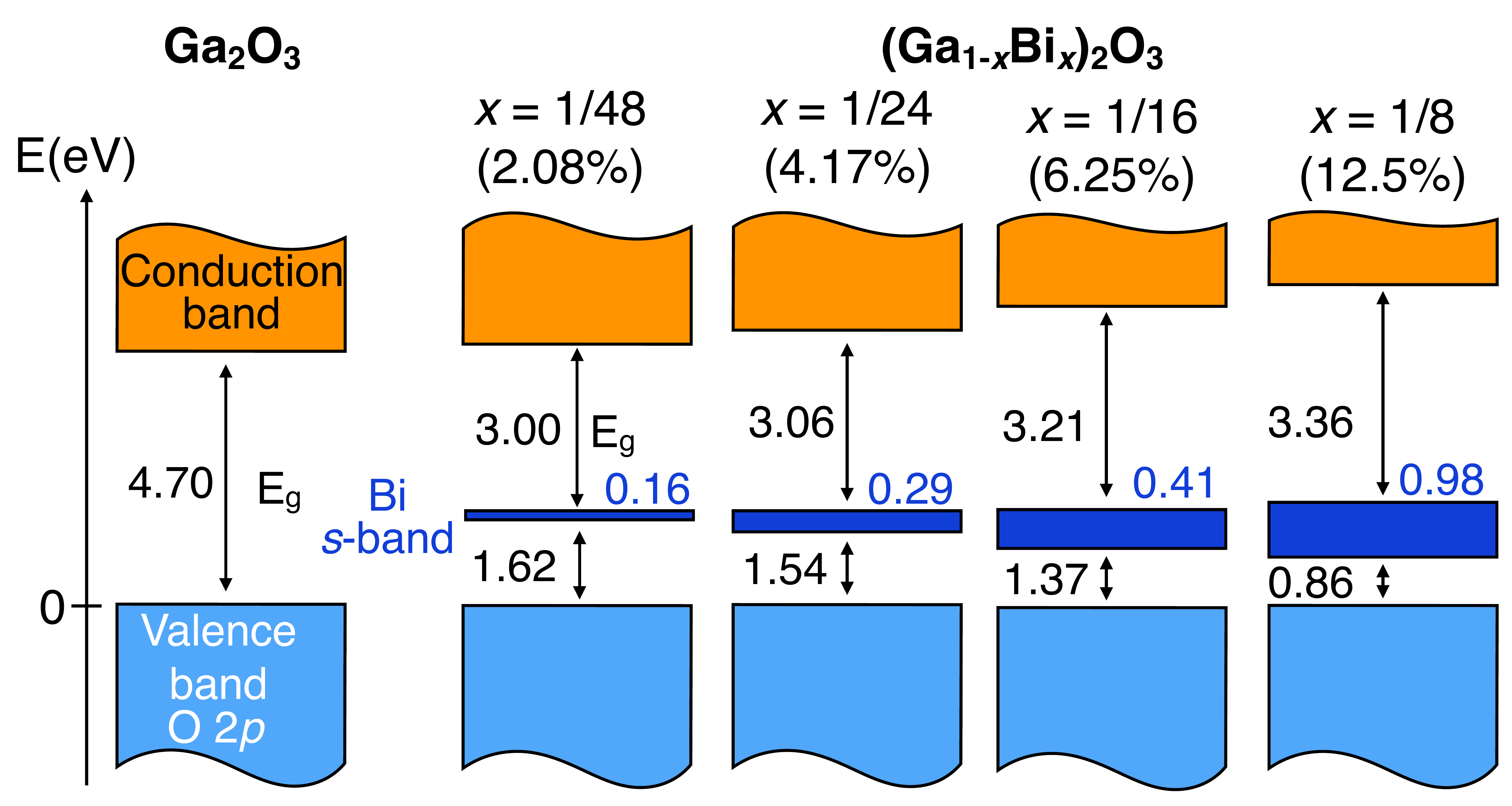}
\caption{
Schematic representation of the effects of adding Bi on the electronic band structure of $\beta$-\ce{Ga2O3}.
\ce{(Ga_{1-$x$}Bi_{$x$})_2O3}  with $x = 1/48$, $1/24$, $1/16$ and $1/8$. All energies are in eV.
The bands showed in dark blue represent the
intermediate valence band, composed of hybridized \ce{Bi} 6$s$ and \ce{O} 2$p$ orbitals.
All the bands are aligned with respect to the \ce{Ga2O3} \ce{O} 2$p$-band, which is used as reference
and placed at the origin of the energy axis.
}
\label{band_alignment}
\end{figure}

The evolution of the band structure of the \ce{(Ga_{1-$x$}Bi_{$x$})_2O3} alloys with increasing
\ce{Bi} concentration is schematically shown in Fig.~\ref{band_alignment}.
Besides the emergence of the intermediate valence band, we also observe an upward shift
of the CBM, which we attribute to a repulsion between the \ce{Bi} intermediate valence band
and the CBM of the host since both have an antibonding character and are composed of $s$ orbitals.
This repulsion increases the band gap of the \ce{(Ga_{1-$x$}Bi_{$x$})_2O3} alloy
as the \ce{Bi} concentration increases. The width of the intermediate valence band also
increases with \ce{Bi} concentration, such that hole localization in the
form of small polarons is expected to be less severe for higher \ce{Bi} content.
Therefore, both the width of the intermediate valence band and the band gap can
be tuned in \ce{(Ga_{1-$x$}Bi_{$x$})_2O3} alloys with \ce{Bi} concentration.

\begin{figure*}[t!]
\centering
\includegraphics[width=0.80\linewidth]{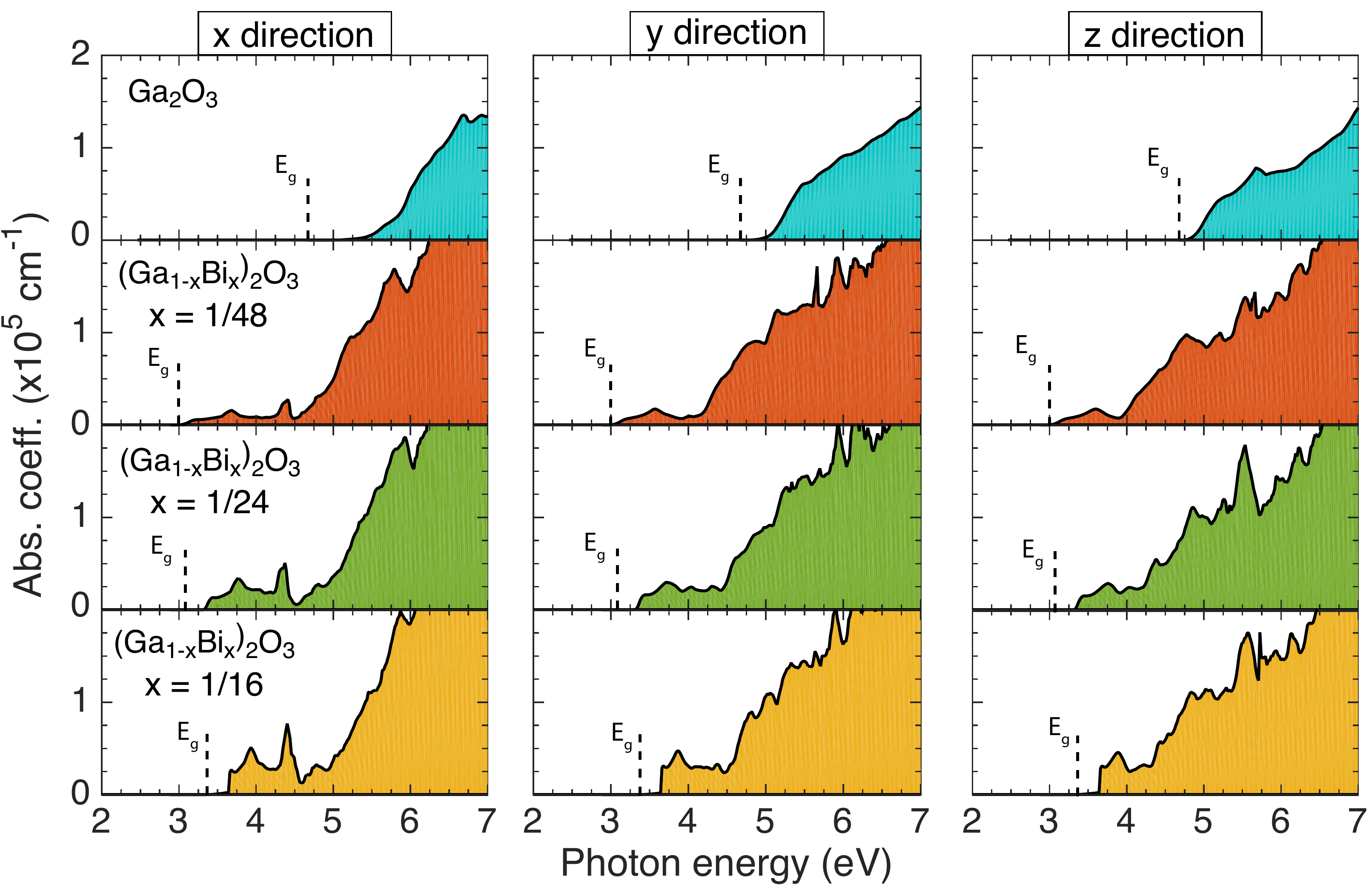}
\caption{
Absorption coefficients as function of the photon energy for: $\beta$-\ce{Ga2O3},
\ce{(Ga_{1-$x$}Bi_{$x$})_2O3} with $x = 1/48$, $1/24$ and $1/16$. The black vertical
dashed line indicate the fundamental.
}
\label{optical}
\end{figure*}

Since the incorporation of even dilute concentrations of \ce{Bi}
significantly changes the electronic structure of \ce{Ga2O3},
we expect that the optical properties of the \ce{(Ga_{1-$x$}Bi_{$x$})_2O3} alloys
will also strongly deviate from that of the parent compound.
This effect can be observed in the calculated absorption coefficient of the alloys
shown in Fig.~\ref{optical}.
Due to the computational cost, we performed these calculations using the PBEsol functional
and applied the scissors operator to correct the fundamental band gap according to the HSE results.

First, we observe that \ce{Ga2O3} shows a sizeable optical anisotropy of the absorption threshold,
which is attributed to the anisotropy of the monoclinic crystal structure \cite{Sabino_205308_2015}. Only
in the z direction the fundamental and optical band gaps coincide. In the x and y directions,
we observe forbidden, or very week transitions near the fundamental band gap, resulting
in optical band gaps that are \SI{0.56}{\electronvolt} and \SI{0.20}{\electronvolt} larger than the
fundamental gap.


The incorporation of \ce{Bi} in \ce{Ga2O3} in the form of dilute \ce{(Ga_{1-$x$}Bi_{$x$})_2O3} alloys
leads to a sizeable red shift of the absorption threshold due to transitions from the intermediate
valence band to the conduction band.  The absorption coefficients near the threshold
in the alloys is much higher than that in the parent compound \ce{Ga2O3}, and increases with \ce{Bi} content
which we attributed to the increased density of states of the intermediate valence band.
We note, however, in the simulations of the random alloys using SQS structures, we
find a disparity between the fundamental gap, labeled $E_g$ in Figure~\ref{optical},
and the absorption threshold due to forbidden or weak transitions near the $\Gamma$ point.
This forbidden or weak transitions is explained by the fact that the intermediate valence band and the
conduction band have the same symmetry at $\Gamma$. In practice, in truly
random alloys, we expect the $k$ dependence to disappear due to the lack of long-range
symmetry and all bands to fold to the zone center, decreasing the disparity observed in our calculations.

We now address the possibility of achieving $p$-type conductivity in the \ce{(Ga_{1-$x$}Bi_{$x$})_2O3}
alloys. Conventional candidate impurities for $p$-type doping in \ce{Ga2O3}, such as \ce{Mg}, \ce{Cd}, \ce{Zn} and \ce{N}
have been predicted to act as deep acceptors with ionization energies higher than
\SI{1}{\electronvolt} \cite{Lyons_05LT02_2018}. Nevertheless,  all the acceptor levels of
the candidate impurities for $p$-type doping are lower than the top of the intermediate valence
band in the dilute \ce{(Ga_{1-$x$}Bi_{$x$})_2O3} alloys. Therefore, if the wavefunctions of the defect levels is decoupled from the host VBM, one may expect that these impurities in the alloy will lead to delocalized holes
in the top of the intermediate valence band. However, we note that the interaction between some of the
impurities and \ce{Bi} could lead to coupling effects and ending up in deep acceptor levels. Further studies, therefore is needed to 
design non-coventional dopants that can lead to shallow defect levels in this system.

In summary, using hybrid density functional calculations we investigate the electronic
structure of dilute \ce{(Ga_{1-$x$}Bi_{$x$})_2O3} alloys. \ce{Bi} introduces a fully
occupied intermediate valence band that is significantly higher in energy than the host
original valence band, provideing the opportunity to achieve $p$-type doping in this system, as long as
the doped acceptor impurity levels are decoupled from the intermediate states. More importantly, even with the intermediate states, the optical absorption is still beyond the visible range; thus, combined with the $p$-type doping opportunity, 
Bi doped \ce{Ga2O3} is a strong candidate for $p$-type transparent material.
Therefore, adding \ce{Bi} to \ce{Ga2O3} will widen the range of potential applications of this wide-band-gap semiconductor.

FPS and AJ were supported by the National Science Foundation Faculty Early Career Development Program under Grant No. DMR-1652994. XC and SHW were supported by the National Nature Science Foundation of China under Grant No. 11634003 and U1530401. This research was also supported by the the Extreme Science and Engineering Discovery Environment supercomputer facility, National Science Foundation grant number ACI-1053575, and the Information Technologies (IT) resources at the University of Delaware, specifically the high performance computing resources.

%


\end{document}